\documentclass[aps,pra,twocolumn,groupedaddress]{revtex4-1}
\usepackage{graphicx}  % needed for figures
\usepackage{subfigure}
\usepackage{dcolumn}   % needed for some tables
\usepackage{bm}        % for math
\usepackage{amssymb}   % for math
\usepackage{comment}
\usepackage{CJK}
\usepackage{color}
\usepackage{amsmath}
\usepackage{mathrsfs}
\usepackage{mathcomp}
\usepackage{textcomp}
\usepackage{dsfont}
\usepackage{braket}
\newtheorem{theorem}{Theorem}

\begin{document}
% Use the \preprint command to place your local institutional report
% number in the upper righthand corner of the title page in preprint mode.
% Multiple \preprint commands are allowed.
% Use the 'preprintnumbers' class option to override journal defaults
% to display numbers if necessary
%\preprint{}

%Title of paper
\title{Lossy Quantum Optical Metrology with Squeezed States}

\author{Xiao-Xiao Zhang$^{1,2}$}
\author{Yu-Xiang Yang$^1$}
\author{Xiang-Bin Wang$^{1,3}$}
\email{xbwang@mail.tsinghua.edu.cn}
\affiliation{$^1$Department of Physics and State Key Laboratory of Low-Dimensional Quantum Physics, Tsinghua University, Beijing 100084, People's Republic of China}
\affiliation{$^2$Department of Applied Physics, University of Tokyo, Tokyo 113-8656, Japan}
\affiliation{$^3$Jinan Institute of Quantum Technology, Shandong Academy of Information and Communication Technology, Jinan 250101, People's Republic of China}

\date{\today}

\newcommand\dd{\mathrm{d}}
\newcommand\ii{\mathrm{i}}
\newcommand\ee{\mathrm{e}}
\makeatletter
\def\ExtendSymbol#1#2#3#4#5{\ext@arrow 0099{\arrowfill@#1#2#3}{#4}{#5}}
\newcommand\LongEqual[2][]{\ExtendSymbol{=}{=}{=}{#1}{#2}}
\makeatother

\begin{abstract}
We study the precise phase estimation using squeezed states in the presence of photon loss. Our exact quantum Fisher information calculation shows significant  enhancement even though the loss is very large and sets a benchmark for experimental realization. We show that if we blindly use the existing parity measurement scheme [P.M. Anisimov \emph{et al.}, Phys. Rev. Lett. \textbf{104}, 103602 (2010)] for the ideal case, the result will be even worse than the classical case given very small loss. Using our formulas, we can optimize the measurement result by making loss-dependent phase shift and choosing appropriate squeezed states (average photon number of squeezed states).
\end{abstract}
% insert suggested PACS numbers in braces on next line
% insert suggested keywords - APS authors don't need to do this
%\keywords{}

%\maketitle must follow title, authors, abstract, \pacs, and \keywords
\maketitle

% body of paper here - Use proper section commands
% References should be done using the \cite, \ref, and \label commands
\section{Introduction}
Many important tasks in scientific research involve physical processes related to phase estimation. It is therefore crucial to study the limit of precision in phase estimation, which is a main task of quantum metrology. By finding measurements that optimally resolve neighboring quantum states, Braunstein and Caves\cite{Braunstein} proposed a metric on density operator space and a generalized uncertainty principle. These lead to an error bound $\delta\phi\ge1/\sqrt{\nu F_Q}$ while estimating a phase $\phi$. Here $F_{Q}$ is called the quantum Fisher information (QFI) of the probe state used and $\nu$ is the number of identical measurements repeated. Employing quantum entanglement, quantum metrology promises higher precision limit in phase estimation\cite{Giovannetti}, the Heisenberg limit (HL) $1/n$, rather than the standard quantum limit (SQL) $1/\sqrt{n}$ ($n$ is the particle number involved).

Imperfections in practical realizations can weaken or destroy advantages of quantum metrology \cite{Dorner, Huelga, Escher, Datta}. Various upper bounds and numerical analysis \cite{Dorner,Escher,Demkowicz} of QFI with noises have been studied. Moreover, the real-world noise can make the measurement task more complicated and intractable as well. For instance, as we find in this paper, if we blindly use the existing parity measurement scheme \cite{Anisimov} for the ideal case, the result is worse than even classical methods given very small losses.

In this paper, we study precise phase estimation using squeezed states with photon loss. We first analyze the entangled two-mode case with exact QFI for lossy channels calculated via the fidelity approach. Much more precise mastery over lossy quantum-optical metrology for squeezed states is therefore achieved, which still shows significant quantum enhancement. We then analyze the loss-dependent result of the parity measurement. We show that predetection phase shift should be done and that intensity of the initial two mode squeezed state (TMSV) should be chosen in order to optimize the result given the channel loss. With our optimization in taking predetection phase shift and choosing intensity of TMSV according to the channel loss, there are still considerable advantages in quantum metrology than in SQL even under lossy channel.

\section{Exact QFI formula with Photon Loss}\label{QFI}
\begin{figure*}
%\begin{center}
  % Requires \usepackage{graphicx}
%  \scalebox{0.5}{\includegraphics*[180,90][610,217]{fig0.eps}}
  \scalebox{0.7}{\includegraphics*[180,110][900,330]{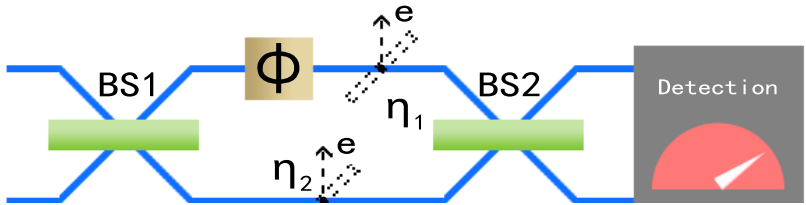}}
  \caption{(Color online) Setup for quantum metrology with MZI. BS1, BS2: 50:50 beam splitters. There are losses for light beams in each path; $\eta_1,\eta_2$ represent transmissivity of light beams. $\phi$ is a phase shifter which takes the amount of phase shift $\pm\phi$ to creation and annihilation operators of the incident light mode.}\label{fig0}
%  \caption{}\label{fig2}
%\end{center}
\end{figure*}

The two-arm interferometric quantum metrology is shown in Fig. \ref{fig0}. Schematically, TMSV light passes through a Mach-Zehnder interferometer(MZI) and measurement is performed at the output ports. The MZI consists of two 50:50 beam splitters and a phase shifter in one arm. The channel loss can be modeled by virtual beam splitters with ancillary modes of the environment. Fictitious beam splitters with transmissivity $\eta_1,\eta_2$ are assumed to produce photon losses here. The loss indeed accompanies phase accumulation simultaneously. But this can be equivalently simulated by separate and commutative phase shifting and photon loss. This can be easily seen from the Kraus operator form of the fictitious beam splitter in the dispersive arm. Without any loss of generality, we can place the phase shifter before the beam splitters.

A Gaussian state can be represented by the covariance matrix\cite{Wang2}. Suppose the covariance matrix of the initial state is $\gamma_{0}$ and the environment (which is vacuum) is $I_e$. The total covariance matrix is $\gamma_0 \oplus I_e$. The state after photon loss and phase shifting is characterized by (see Theorem \ref{density_matrix} in Appendix \ref{append0})
\begin{equation}\label{equ00}
\gamma =  M_{loss}M_\phi(\gamma_{0}\oplus I_e)M_\phi^{T}M_{loss}^{T}=\left(\begin{array}{cc}
\gamma' & \gamma_{S}\\
\gamma_{S}^{T} & \gamma_{E}
\end{array}\right),
\end{equation}
where $M_{loss/\phi}$ represents the linear transformation matrix of field operators caused by the virtual beam splitters or the phase shifter, in particular,
 $M_{loss/\phi}=(\oplus_{j=1}^4 K)M_{loss/\phi}'(\oplus_{j=1}^4 K^{-1})$,
 where $K=\left(
\begin{array}{cc}
1 & 1 \\
\ii & -\ii \\
\end{array}
\right)/\sqrt{2}
$. Matrix $M_{loss}'$ corresponds to the transformations $\hat{a}_{i}\rightarrow \sqrt{\eta_i}\hat{a}_{i}-\sqrt{1-\eta_i}\hat{e}_{i} \,, \hat{e}_{i}\rightarrow \sqrt{1-\eta_i}\hat{a}_{i}+\sqrt{\eta_i}\hat{e}_{i}$ and their Hermite conjugations, where notations $\hat{a}_i\,,\hat{e}_i$ are the $i$th ($i=1,2$) arm's and its environment mode's annihilation operators, respectively. Similarly, matrix $M_{\phi}'$ corresponds to transformations $\hat{a}_{1}\rightarrow \ee^{-\ii \phi}\hat{a}_1$ and its Hermite conjugation.
 Tracing the environment, the covariance matrix of the reduced density operator for the system we considered is just $\gamma'=\gamma'(\phi)$.

If the initial state is a TMSV, i.e.,
\[\gamma_0=\left(
\begin{array}{cccc}
\cosh r & 0 & \sinh r & 0 \\
0 & \cosh r & 0 & -\sinh r \\
\sinh r & 0 & \cosh r & 0 \\
0 & -\sinh r & 0 & \cosh r \\
\end{array}
\right),
\]
the final state can be calculated using (\ref{equ00}). And it is characterized by
\begin{equation}\label{equ12}
\gamma'(\phi)=\left(\begin{array}{cccc}
d_1 & 0 & a\cos{\phi} & -a\sin{\phi}\\
0 & d_1 & -a\sin{\phi} & -a\cos{\phi}\\
a\cos{\phi} & -a\sin{\phi} & d_2 & 0\\
-a\sin{\phi} & -a\cos{\phi} & 0 & d_2
\end{array}\right),
\end{equation}
where $d_1=1+2\eta_1\sinh^2{r}$, $d_2=1+2\eta_2\sinh^2{r}$, $a=\sqrt{\eta_1\eta_2}\sinh{2r}$ and $r$ is the squeeze parameter.

From Eq. \ref{equ12} we can calculate the QFI for lossy channels with the method provided in \cite{Paraoanu,Wang1} (see Appendix \ref{append0}). The result turns out to be
\begin{equation}\label{equ13}
F_{Q}=\frac{2n(n+2)\eta_1\eta_2}{2+n(\eta_1+\eta_2-2\eta_1\eta_2)},
\end{equation}
where $n=2\sinh^2{r}$ is the average photon number of the initial ideal TMSV state. Obviously, when $\eta_1=\eta_2=1$, this is reduced to the ideal case, $F_Q=n(n+2)$. We shall call it ``modified HL.''  For equal losses in both arms ($\eta_1=\eta_2=\eta$) and one-arm losses ($\eta_1=\eta,\eta_2=1$), the QFIs are
$\frac{n(n+2)\eta^2}{1+n(1-\eta)\eta}$ and $\frac{2\eta n(n+2)}{n(1-\eta)+2}$, respectively. The latter is more practical since losses are mostly generated during phase shift operation.
\begin{figure*}
%\begin{center}
  % Requires \usepackage{graphicx}
  \includegraphics[width=0.40\linewidth]{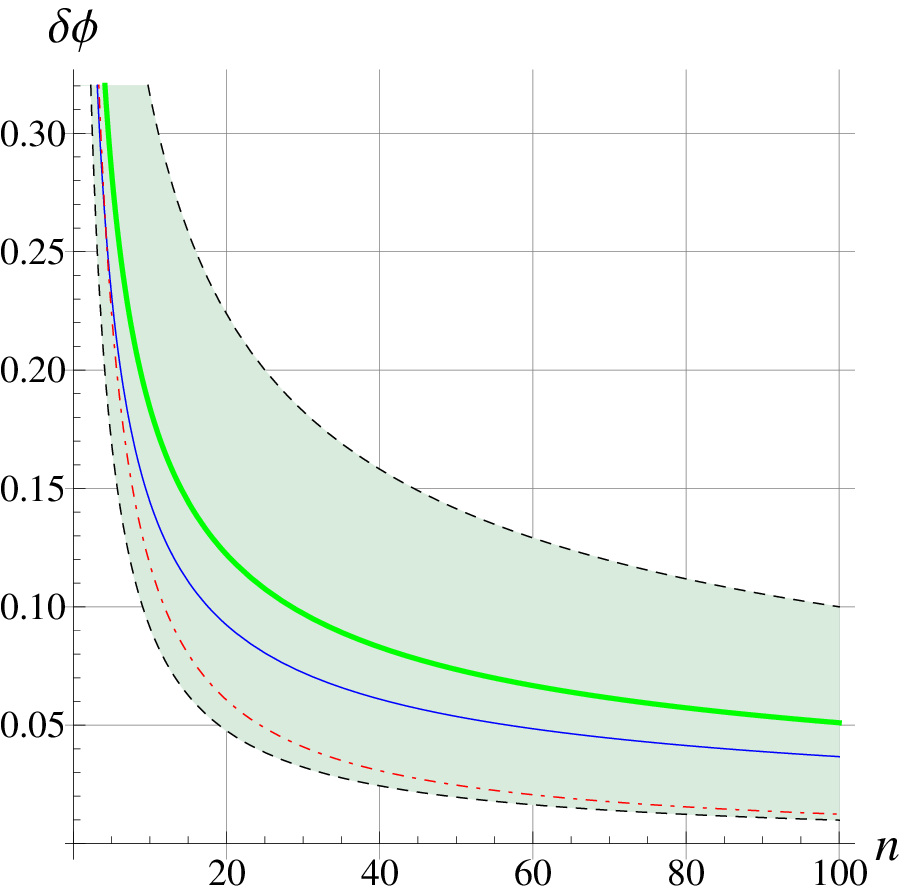}
  \includegraphics[width=0.40\linewidth]{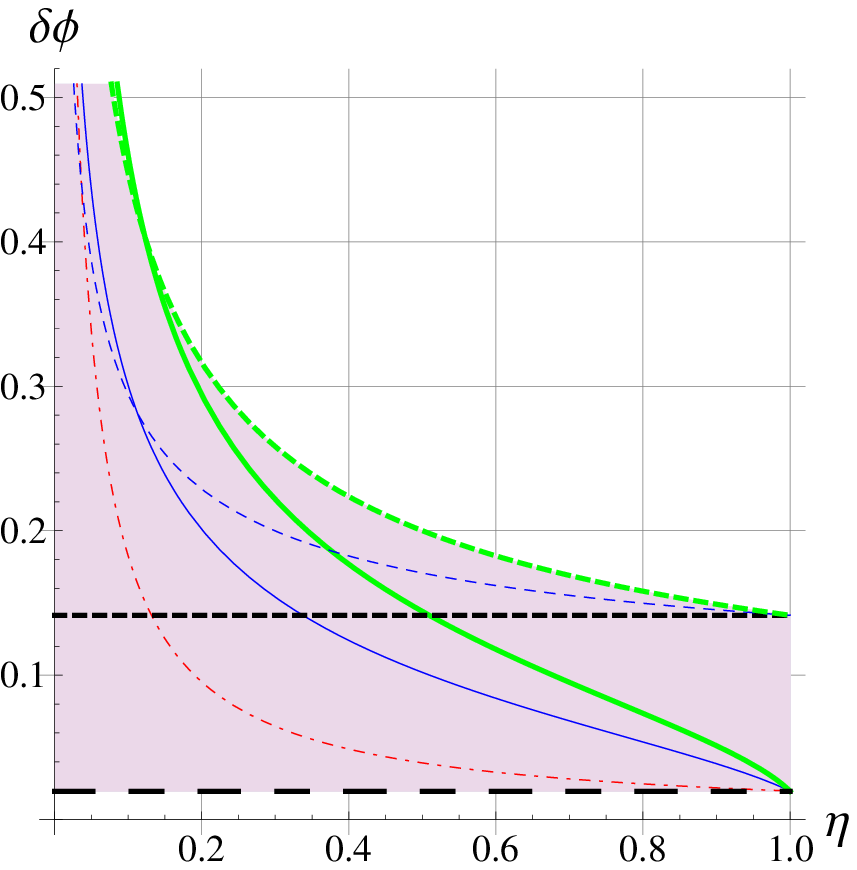}
  \caption{(Color online) (Left) Estimation precision $\delta\phi$ vs average photon number $n$. Lowest (dashed black) line: modified HL. Uppermost (dashed black) line: SQL. Thick solid green line and thin solid blue line: quantum limits $\frac{1}{\sqrt{F_Q}}$ for two-arm-equal-loss($\eta_1=\eta_2=0.8$) and one-arm-loss($\eta_1=0.8,\eta_2=1$) models, respectively. Dot-dashed red line: \cite{upper bound}. (Right) Estimation precision $\delta\phi$ v.s. transmissivity $\eta$. Lower horizontal dashed black line: modified HL. Upper horizontal dashed black line: SQL. Thick solid green line and thin solid blue line: quantum limits $\frac{1}{\sqrt{F_Q}}$ for two-arm-equal-loss and one-arm-loss models, respectively. Thick dashed green line and thin dashed blue line: classical limits \cite{Dorner} $\frac{1}{\sqrt{n\eta}}$ for two-arm-equal-loss and $\frac{1+\sqrt{\eta}}{2\sqrt{n\eta}}$ for one-arm-loss models, respectively. Dot-dashed red line: \cite{upper bound}.}
  \label{fig12}
%  \caption{}\label{fig2}
%\end{center}
\end{figure*}

Figure \ref{fig12} shows the comparison between (\ref{equ13}) and several different characterizations of the usefulness of states. See the caption under Fig. \ref{fig12} for detailed description. Here we can see our exact formula of QFI (\ref{equ13}) (corresponds to ``quantum limit'' $1/\sqrt{F_Q}$ in Fig. \ref{fig12}) for lossy scenarios of the above two particular cases is in agreement with the general upper bounds in \cite{Escher}. The general upper bound \cite{Escher} of QFI for optical interferometry gives scaling $1/\sqrt{n}$ for $n\gg\eta/(1-\eta)$ and $1/n$ for $n\ll\eta/(1-\eta)$. Despite the same physical model, the intuitive upper bound of QFI \cite{upper bound} generally acts worse than our exact QFIs as depicted. This implies the importance of our exact QFI study.
As for the performance of the TMSV state, the QFIs of the two lossy scenarios approach their classical limits in very large loss limits($\eta\rightarrow0$).
 Notably, our formula shows that the TMSV state beats the SQL even in the presence of large losses ($\eta=0.6$ for two-arm losses, $\eta=0.4$ for one-arm losses). This implies that squeezed states together with a cleverly designed measurement scheme can show quantum enhancement under a very lossy channel.

\section{Measurement Scheme}\label{scheme}
Our purpose here is to show a useful measurement scheme to exploit the quantum advantage described in Sec. \ref{QFI}. In spite of the fact that QFI very well quantifies the potential advantage of a certain state in quantum metrology, note that not every detection method is able to achieve this advantage, especially in noisy scenarios. In terms of our probe state and evolution setup, similar to \cite{Anisimov}, we make use of the so-called parity measurement, which can be applied to precise phase estimation for a wide range of photonic states \cite{Gerry}. In fact, photon loss may occur both inside the interferometer and at the measurement stage. The effect of photons lost in detection varies with the specific realization of parity measurement. And loss inside the interferometer affects the QFI, i.e., the actual potence of a state in quantum metrology. Therefore, we deal only with photon loss inside the interferometer in order to comprehend the advantage of the TMSV state and a possible suitable measurement scheme.

In our scheme, we need the parity measurement $\hat{\Pi} = \exp(-\ii\pi \hat{a}_1^\dag \hat{a}_1)$ (we specify $a_1$ as the mode we measure). Interestingly, highly reliable photon number resolving detectors are not necessary for parity measurement. Actually, parity measurement can be done without photon number resolving detectors \cite{Wigner0,Plick}. The value of the Wigner function at the origin is the same with the parity of one mode field \cite{Wigner0}: $\braket{\hat{\Pi}}=\frac{\pi}{2}W(0,0)$. Quantum tomography techniques provide a means to reconstruct the Wigner function of a radiation field. Since $W(0,0)$ is the only thing we need for the phase estimation, one can use balanced homodyne detection \cite{Plick} to demonstrate our scheme.  Detailed discussion about the setups and mathematical derivations are given in \cite{Plick}.
\begin{figure*}
%\begin{center}
  % Requires \usepackage{graphicx}
  \includegraphics[width=0.42\linewidth]{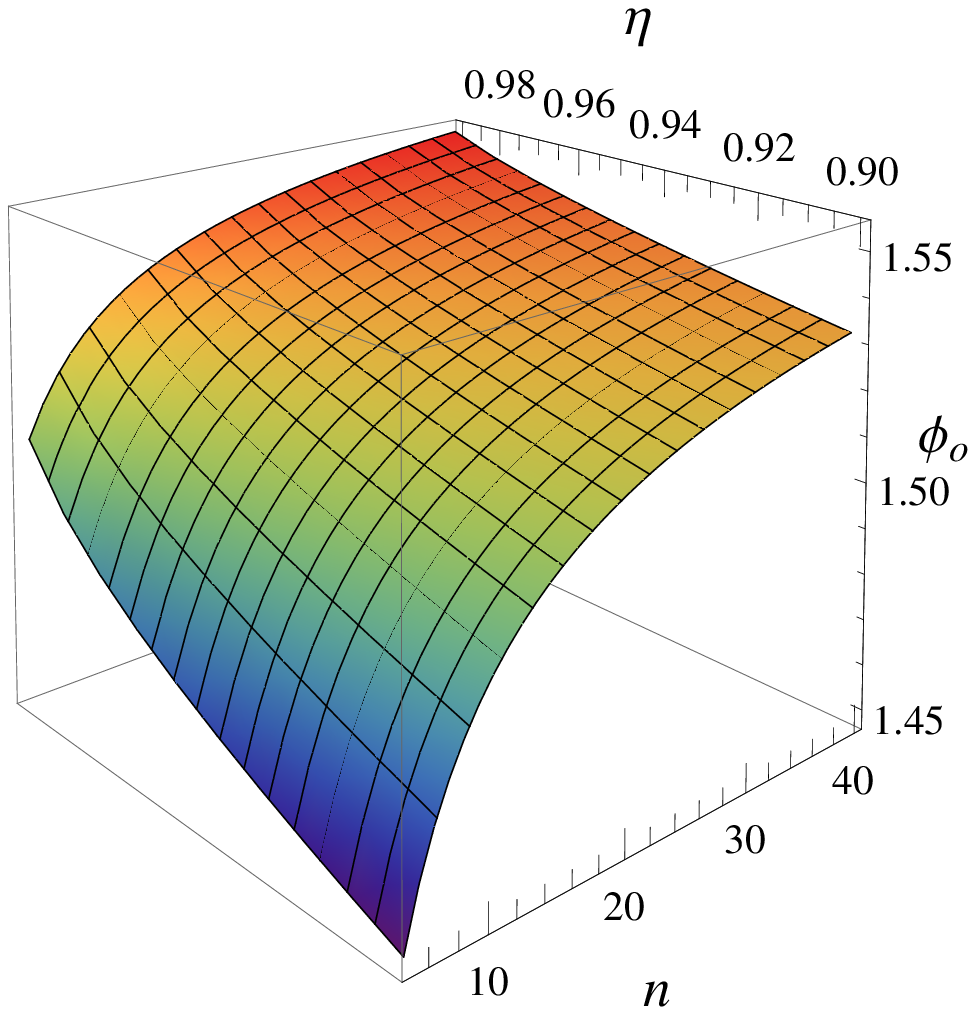}
  \includegraphics[width=0.38\linewidth]{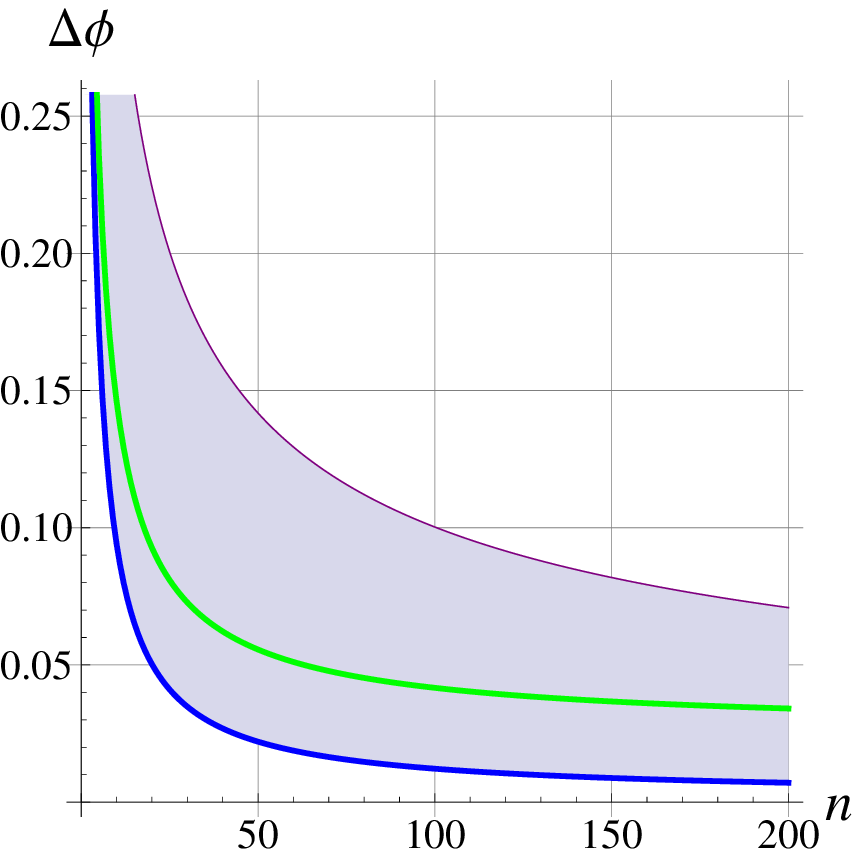}
  \caption{(Color online) (Left) Optimal measurement point $\phi_o$ vs average photon number $n$ and transmissivity $\eta$. %\hfill\\ %\hspace*{9mm}
  (Right) Phase estimation error $\Delta\phi$ vs average photon number $n$ for one-arm losses ($\eta=99\%$). Purple (uppermost) line: classical limit \cite{Dorner}. Blue (lowest) line: quantum limit of exact QFI. Green (middle) line: qptimal phase measurement error via parity detection.}\label{fig34}
%\end{center}
\end{figure*}

We denote the state evolution in MZI as $\ket{\Psi_f}=\hat{U}_{lossyMZI}\ket{\psi_{TMSV}}_{a_1,a_2}\otimes\ket{0,0}_{e_1,e_2}$, where $\hat{U}_{lossyMZI}=\hat{U}_{BS}\hat{U}_{loss}\hat{U}_{\phi}\hat{U}_{BS}$. So the expectation value of parity measurement at the output mode is $\braket{\hat{\Pi}}_{\phi}=\braket{\Psi_f|\hat{\Pi}|\Psi_f}$. Using an exponential operator reordering method \cite{Wang3} (see Appendix \ref{append1}) henceforward, we can calculate the $\hat{U}_{lossyMZI}$ operator and expectation values of observables. Thus, we obtain
\begin{equation}\label{equ14}
\braket{\hat{\Pi}}_{\phi}=\sqrt{\frac{2}{n\omega+2}},
\end{equation}
where $\omega=2\eta_1\eta_2(n+2)\cos^2{\phi}+(\eta_1+\eta_2)(2-\eta_1-\eta_2)$.
Explicitly, the variance of phase estimation quantifying quantum enhancement is
\begin{equation}\label{equ15}
\Delta\phi^2=\frac{\braket{\hat{\Pi}^2}_{\phi}-\braket{\hat{\Pi}}_{\phi}^2}{\left(\partial \braket{\hat{\Pi}}_{\phi} / \partial  \phi\right)^2}=\frac{\omega(n\omega+2)^2}{2\eta_1^2\eta_2^2n(n+2)^2\sin^2{2\phi}}.
\end{equation}
Equations (\ref{equ14}) and (\ref{equ15}) reduce to the ideal case in \cite{Anisimov} when $\eta_1=\eta_2=1$: $\widetilde{\braket{\hat{\Pi}}_{\phi}}=\frac{1}{\sqrt{1+n(n+2)\cos^2{\phi}}},\widetilde{\Delta\phi}=\frac{1+n(n+2)\cos^2{\phi}}{|\sin{\phi}|\sqrt{n(n+2)}}$.
This ideal $\widetilde{\Delta\phi}$ is monotonic to $\phi$ if $\phi\in[0,\pi/2]$, i.e. the best precision is achieved when $\phi=\pi/2$. When there are photon losses, our Eq. (\ref{equ15}) shows that $\phi$ is no longer monotonic with respect to $\phi$. Actually, in such a case, the relationship $\Delta\phi=\Delta\phi(n,\eta_1,\eta_2,\phi)$ is rather complicated. (1) It depends on the values of $n,\eta_1,\eta_2$.  (2) Equation (\ref{equ15}) can give us an optimal point phase value $\phi_o=\phi_o(n,\eta_1,\eta_2)$ numerically. This means that, if the unknown phase shift is small, or, if we have the knowledge of a small range for the unknown phase shift, then we can first take a suitable phase shift according to the known values of $n,\eta_1,\eta_2$ so as to achieve the highest precision in given conditions. That is to say, instead of exploiting quantum enhancement around $\phi=\pi/2$, we can attain the minimal $\Delta\phi$ if we adjust the phase to be near to this $\phi_o$ and then measure it. (See the next paragraph for details.) (3) Without this ``optimal measurement point'' amendment, the result of phase estimation through ordinary parity measurement will be poor; normally, no enhancement can be achieved. [This is  not shown in Fig. \ref{fig34} (Right) since it behaves poorly.]

The above discussion shows that, in practical experiment settings with photon losses, we have to make a biased measurement in order to achieve the best precision. Say, we first calculate the optimal phase $\phi_o$ by Eq. (\ref{equ15}), given $n,\eta_1,\eta_2$. If the phase $\phi$ to be measured is small, we then add an additional phase shift of $\varphi=\phi_o$. More generally, if the unknown phase $\phi$ is in a small range centered at $\tilde{\phi}$, we shall add an additional phase shift $\varphi = \phi_o -\tilde{\phi}$. After parity measurement, we obtain $\varphi$ and then deduce $\phi$. As can be easily shown, the precision of $\phi$ is equal to that of $\varphi$. Therefore, by taking an additional phase shift, we can achieve a precise result of $\phi$ even though there are channel losses.

\begin{figure}
\subfigure[]{
\label{fig:mini:subfig:a} %% label for first subfigure
\begin{minipage}[c]{0.25\textwidth}
\centering
\includegraphics[width=0.66\linewidth]{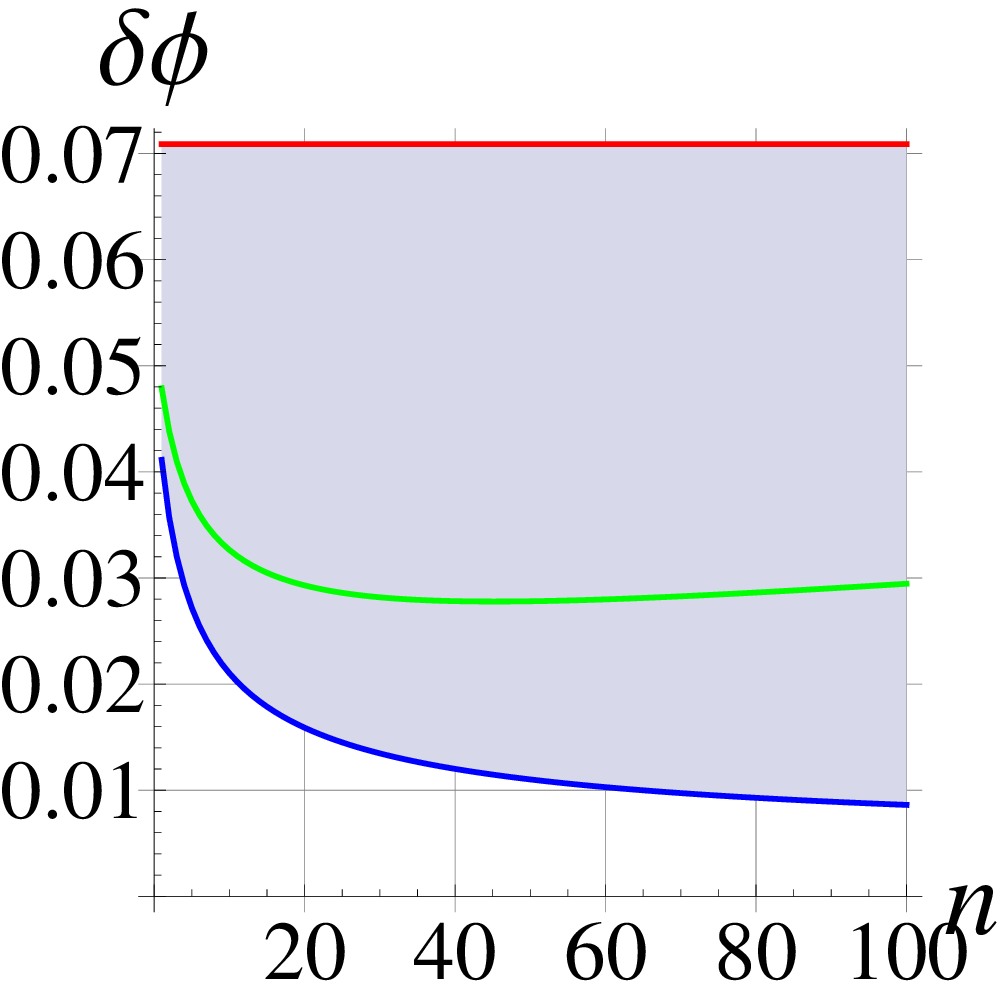}
\end{minipage}}%
\subfigure[]{
\label{fig:mini:subfig:b} %% label for first subfigure
\begin{minipage}[c]{0.25\textwidth}
\centering
\includegraphics[width=0.66\linewidth]{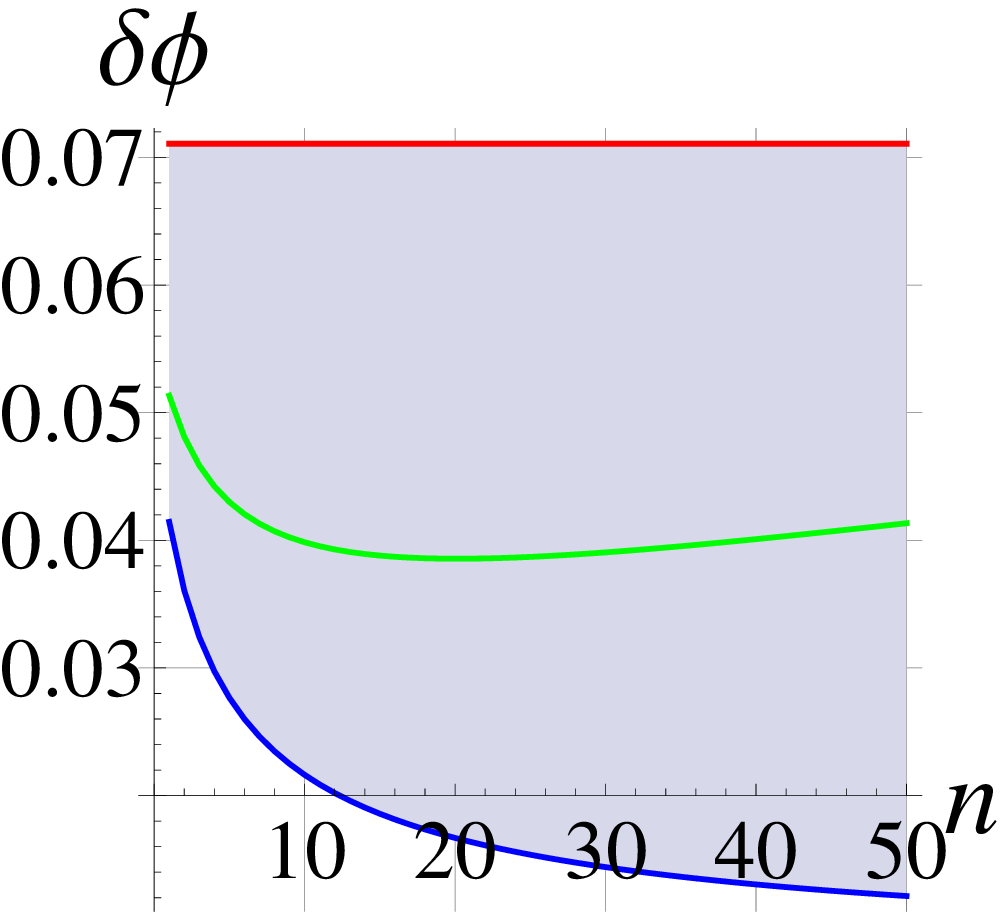}
\end{minipage}}%
\hfill\\
\subfigure[]{
\label{fig:mini:subfig:c} %% label for first subfigure
\begin{minipage}[c]{0.25\textwidth}
\centering
\includegraphics[width=0.66\linewidth]{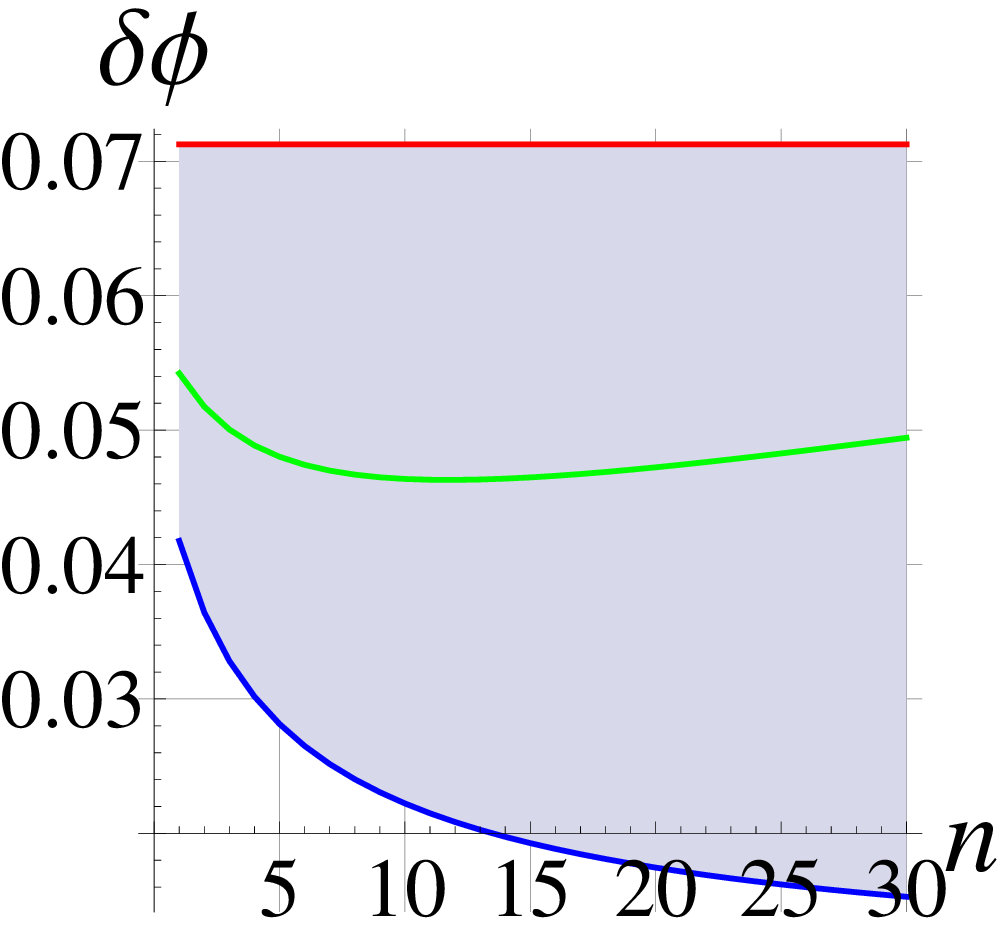}
\end{minipage}}%
\subfigure[]{
\label{fig:mini:subfig:d} %% label for second subfigure
\begin{minipage}[c]{0.25\textwidth}
\centering
\includegraphics[width=0.66\linewidth]{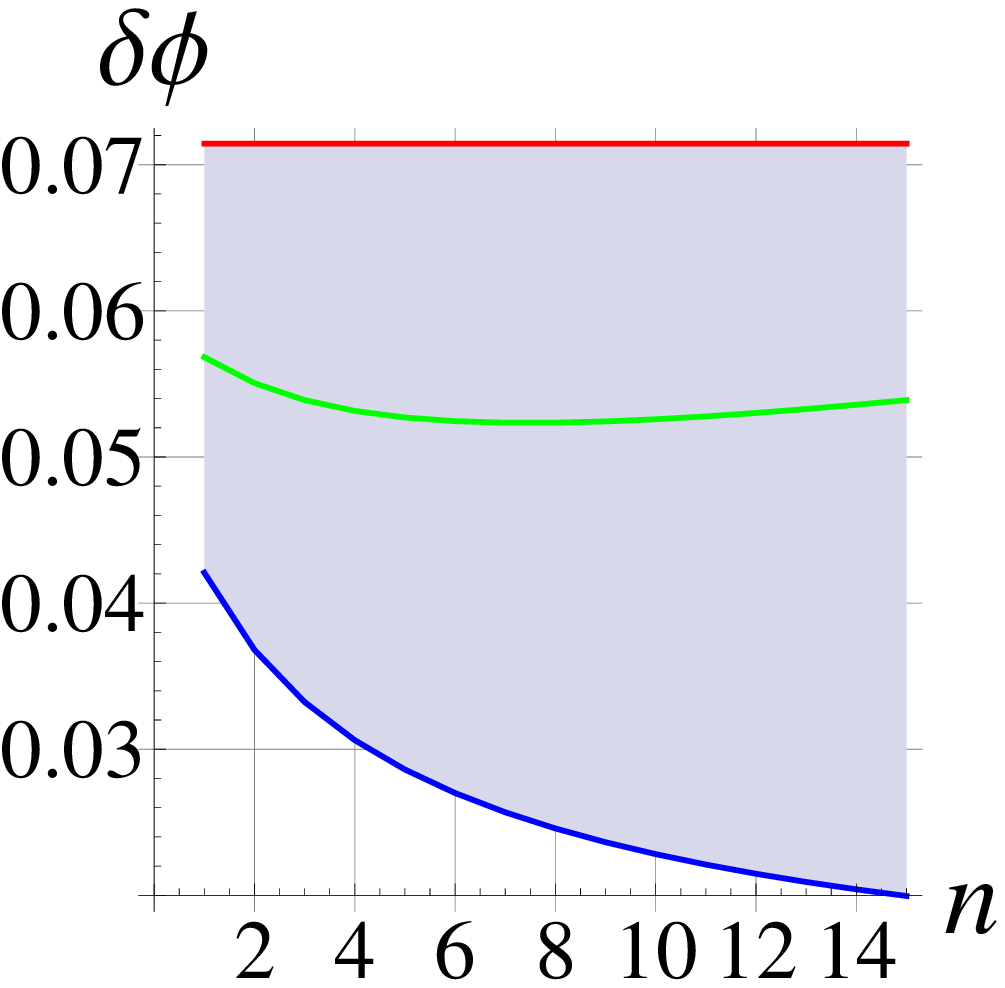}
\end{minipage}}
\caption{(Color online) Repeated-experiment phase error $\delta\phi$ vs average photon number $n$ for one-arm losses [$\eta=99\%,98\%,97\%,96\%$ from (a) to (d), $N=200$ fixed]. Red horizontal line: classical limit \cite{Dorner}. Blue (lowest) line: quantum limit of exact QFI. Green (middle) line: optimal phase measurement error via parity detection.}
\label{fig5678}%\label{fig:mini:subfig} %% label for entire figure
\end{figure}

Since losses are mostly generated by phase shifting, we focus on the case of $\eta_1=\eta,\eta_2=1$ henceforward. Interestingly, now this $\phi_o$ is monotonic to $n$ and $\eta$ separately, as shown in Fig. \ref{fig34} (Left).
The performance of parity measurement at an optimal measurement point with losses in the dispersive arm has been calculated. Compared with our exact lossy QFI and the classical limit $\frac{1+\sqrt{\eta}}{2\sqrt{n\eta}}$ for coherent state \cite{Dorner}, significant quantum enhancement is achieved in the $\eta=0.99$ case as shown in Fig. \ref{fig34} (Right). To our knowledge of contemporary optical technology in laboratories, this is doable for a proof-of-principle demonstration. In addition, our lossy QFI bound in Sec. \ref{QFI} cannot be totally achieved through our parity measurement scheme. Fixing total photon number $N=200$ (see the next paragraph), we also calculate the performance of the scheme for cases of $\eta=0.99, 0.98,0.97,0.96$ as shown in Fig. \ref{fig5678}.

%\begin{figure}
%%\begin{center}
%  % Requires \usepackage{graphicx}
%  \includegraphics[width=0.33\linewidth]{fig5.eps}
%  \includegraphics[width=0.36\linewidth]{fig6.eps}
%  \includegraphics[width=0.35\linewidth]{fig7.eps}
%  \includegraphics[width=0.34\linewidth]{fig8.eps}
%  \caption{(Color online) Repeated-experiment phase error $\delta\phi$ v.s. average photon number $n$ for one-arm losses ($\eta=99\%,98\%,97\%,96\%$ from left to right, $N=200$ fixed). Red horizontal line: Classical limit\cite{Dorner}. Blue (lowest) line: Quantum limit of exact QFI. Green (middle) line: Optimal phase measurement error via parity detection.}\label{fig5678}
%%\end{center}
%\end{figure}

As for the situation closer to real experimental setups, we in general repeat an identical experiment many times to reduce error classically. So we divide our $N$-photon resource into $\nu$ identical states. $n=N/\nu$ is just interpreted as the average photon number in the above discussion. Therefore the more general formula of phase estimation error is $\delta\phi(n,\eta,\phi)=\frac{\Delta\phi(n,\eta,\phi)}{\sqrt{N/n}}$. With $N$ fixed, the optimization of phase measurement point is similar to (\ref{equ15}). However, as shown in Fig. \ref{fig5678}, when $\eta$ and $N$ are fixed, there turns out to be an optimal average photon number $n_o$ that increases with $\eta$ to attain minimal error $\delta\phi$. So the averaged photon number in the initial TMSV state is not the larger the better, which means, interestingly, that the squeezing is not the higher the better in order to get the largest quantum enhancement in phase estimation under lossy channel. There exists an optimal value for squeezing given total photon resource and channel loss.

Calculation shows that the measurement scheme here will almost completely lose its quantum advantage when $\eta$ goes to $90\%$. This means that the scheme presented in Sec. \ref{scheme} does not fully realize the quantum enhancement shown by QFI calculation in Sec. \ref{QFI}. Note that this does not conflict with the result of largest loss given in Sec. \ref{QFI}. QFI provides the {\em mathematically allowed result} for quantum metrology. The scheme used here is a demonstration of an experimentally achievable quantum advantage rather than the one mathematically derived within the QFI framework\cite{Braunstein}, which seems to be far beyond currently existing technologies.

\section{Concluding remark}
In summary, we have presented an explicit formula for quantum Fisher information for phase estimation with a two-mode squeezed state under lossy channels. We presented detailed results on how to optimize the quantum enhancement in phase estimation given channel loss. The measurement scheme in Sec. \ref{scheme} does not fully reach the allowed quantum metrology advantage as calculated by QFI in Sec. \ref{QFI}. This leaves an interesting problem for future study of further improving the quantum enhancement through seeking new feasible measurement schemes.

\section*{Acknowledgments}
X.B.W. proposed this work. X.X.Z. and Y.X.Y. contributed equally to this work. The authors thank Zong-Wen Yu for useful discussion and Bo-Sen Lv for technical assistance. This work was supported in part by the National High-Tech Program of China through
Grants No. 2011AA010800 and No. 2011AA010803 and by NSFC Grants No. 11174177 and No. 60725416.

\appendix
\section{Proof of Eq.\eqref{equ13}: exact QFI for lossy channels}\label{append0}
The calculation of quantum Fisher information relies on the relation between the Bures distance and QFI \cite{Braunstein}:
\begin{equation}
F_{Q}=4[\dd s_{Bures}(\hat{\rho}_{\phi},\hat{\rho}_{\phi+\dd\phi})]^{2}/\dd\phi^{2},
\end{equation}
where $\dd s_{Bures}(\hat{\rho}_{\phi}, \hat{\rho}_{\phi+\dd\phi})=\sqrt{2[1-F(\hat{\rho}_{\phi}, \hat{\rho}_{\phi+\dd\phi})]}$.

Therefore, in order to calculate the QFI we need the expression for the Bures fidelity for two Gaussian states, which has been given in \cite{Wang1} as the main result. Specifically, it can be expressed as the following theorem in our two-mode case:
\begin{theorem}\label{theorem_fidelity}
For two-mode states expressed in the form
\begin{equation}\label{eq:rho-N}
\hat{\rho}_\phi=\exp[-\frac{1}{2}\alpha^TN_\phi\alpha]
\end{equation}
where $\alpha=(\hat{a}_1^\dagger,\hat{a}_1,\hat{a}_2^\dagger,\hat{a}_2)^T$, and $N_\phi$ is a $4\times 4$ symmetric matrix whose elements can be dependent on $\phi$, the Bures fidelity between two states can be expressed as
\begin{widetext}
\begin{equation}\label{Bfidelity}
F(\hat{\rho}_{\phi}, \hat{\rho}_{\phi+\dd\phi})=\frac{|\operatorname{det}(\ee^{-N_\phi\Sigma^{-1}}-I)\operatorname{det}
(\ee^{-N_{\phi+\dd\phi}\Sigma^{-1}}-I)|^{\frac{1}{2}}}{\operatorname{det}[\sqrt{\ee^{-N_\phi\Sigma^{-1}/2}
e^{-N_{\phi+\dd\phi}\Sigma^{-1}}\ee^{-N_\phi\Sigma^{-1}/2}}-I]},
\end{equation}
\end{widetext}
where
\begin{equation*}
\Sigma=\left(\begin{array}{cccc}
0 & 1 & 0 & 0\\
-1 & 0 & 0 & 0\\
0 & 0 & 0 & 1\\
0 & 0 & -1 & 0
\end{array}\right).
\end{equation*}
\end{theorem}
Therefore, to calculate the Bures fidelity, the only thing we need to derive is the expression for $N_\phi$. In the following, $N_\phi$ will be derived from the covariance matrix $\gamma'(\phi)$ of the final state. To make it concise, we shall use the following theorem\cite{Wang2}:
\begin{theorem}\label{density_matrix}
Define $\Gamma=(\hat{x}_1, \hat{p}_1, \cdots ,\hat{x}_n, \hat{p}_n)^T$.
If the density operator for covariance matrix $\gamma_0$ is $\exp[-\frac{1}{2}\Gamma^TN_0\Gamma]$, then the density operator
for covariance matrix $\gamma = M^T \gamma_0 M$ is $\rho = \exp[-\frac{1}{2}\Gamma^T M^TN_0{M^{-1}}^{T}\Gamma] $, given $M$ a complex symplectic matrix.
\end{theorem}
{\em Proof.} Suppose $\rho=\hat S\rho_0 \hat S^\dagger$. Then the characteristic function for $\rho$ is
${\rm tr} [\exp(i\Gamma^T \xi )\hat S\rho_0\hat S^\dagger] ={\rm tr} [\hat S^\dagger \exp(i\Gamma^T \xi )\hat S\rho_0]
= {\rm tr} [\exp(i   \Gamma^T M(\hat S^\dagger)\xi )\rho_0]  $, where $\xi =(\xi_1,\xi_2,\cdots,\xi_{2n})$ and $\hat S^\dagger \Gamma^T \hat S =\Gamma^T M(\hat S^\dagger)$. Hence the covariance matrix for $\rho$ is $M^T(\hat S^\dagger) \gamma_0 M(\hat S^\dagger)$.
On the other hand, we have already assumed  $\gamma = M^T \gamma_0 M$; this shows that $M(\hat S^\dagger)=M$.
Also, we have assumed the form of $\rho=\hat S\rho_0\hat S^\dagger$ and $\rho_0=\exp[-\frac{1}{2}\Gamma^TN_0\Gamma]$. This gives rise to
$\rho = \exp[-\frac{1}{2}\Gamma^TM(\hat S)N_0 M^T (\hat S) \Gamma] $ if $\hat S \Gamma^T \hat S^\dagger = \Gamma^T M(\hat S)$. The uitarity condition
$\hat S\hat S^\dagger =1$ requests $M(\hat S) = M^{-1}(\hat S^\dagger)= M^{-1}$. Therefore $\rho = \exp[-\frac{1}{2}\Gamma^T M^{-1}N_0 {M^{-1}}^T  \Gamma] $. This completes the proof.

In our case,  covariance matrix of the final state is $\gamma'(\phi)$ as given in Eq. (\ref{equ12}).
Diagonalizing matrix $\gamma'(\phi)$, we have the following form for $\gamma'(\phi)$:
\begin{equation}
\gamma'(\phi)=M^TDM,
\end{equation}
where
$D=\operatorname{diag}(r_1,r_1,r_2,r_2)$ with $r_1=\frac{1}{2}(d_1+d_2)\sqrt{4a^2+1}+\frac{1}{2}(d_1-d_2)$ and $r_2=\frac{1}{2}(d_1+d_2)\sqrt{4a^2+1}-\frac{1}{2}(d_1-d_2)$, and
\begin{widetext}
\begin{equation*}
M_\phi=\left(\begin{array}{cccc}
\cosh{r_0}\sin{\frac{\phi}{2}} & -\cosh{r_0}\cos{\frac{\phi}{2}} & \sinh{r_0}\sin{\frac{\phi}{2}} & \sinh{r_0}\cos{\frac{\phi}{2}}\\
\cosh{r_0}\sin{\frac{\phi}{2}} & \sinh{r_0}\cos{\frac{\phi}{2}} & \sinh{r_0}\cos{\frac{\phi}{2}} & -\sinh{r_0}\sin{\frac{\phi}{2}}\\
\sinh{r_0}\sin{\frac{\phi}{2}} & \sinh{r_0}\cos{\frac{\phi}{2}} & \cosh{r_0}\sin{\frac{\phi}{2}} & -\cosh{r_0}\cos{\frac{\phi}{2}}\\
\sinh{r_0}\cos{\frac{\phi}{2}} & -\sinh{r_0}\sin{\frac{\phi}{2}} & \cosh{r_0}\cos{\frac{\phi}{2}} & \cosh{r_0}\sin{\frac{\phi}{2}}
\end{array}\right)
\end{equation*}
\end{widetext}
is symplectic, where $r_0$ satisfies $\coth{4r_0}=-\frac{d_1+d_2}{4a}$
. Define $\hat{\rho}_D$ as the density operator whose covariance matrix is the diagonal matrix $D$.
It is well known that the density operator for diagonal covariance matrix $D$ is a thermal state as
$\hat{\rho}_D=e^{-\frac{1}{2}\alpha^T N_0 \alpha}$ with
\begin{equation*}
N_0=\left(\begin{array}{cccc}
0 & -\operatorname{ln}(\frac{r_1-1}{r_1+1}) & 0 & 0\\
-\operatorname{ln}(\frac{r_1-1}{r_1+1}) & 0 & 0 & 0\\
0 & 0 & 0 & -\operatorname{ln}(\frac{r_2-1}{r_2+1})\\
0 & 0 & -\operatorname{ln}(\frac{r_2-1}{r_2+1}) & 0
\end{array}\right).
\end{equation*}
Therefore the density operator for the final state $\rho$ is
\begin{equation*}\label{eq:rho}
\begin{split}
\hat{\rho}&=\hat{S}\hat{\rho}_D \hat{S}^\dagger=\hat{S}\ee^{-\frac{1}{2}\alpha^T N_0 \alpha }\hat{S}^{\dagger}\\
&=\hat{S}\ee^{-\frac{1}{2}\Gamma^TK N_0 K^T \Gamma}\hat{S}^{\dagger}\\
&=\ee^{-\frac{1}{2}\Gamma^T {M^{-1}} K N_0 K^T {M^{-1}}^T \Gamma}.
\end{split}
\end{equation*}
Here
\begin{equation*}
K=\frac{1}{\sqrt{2}}\left(\begin{array}{cccc}
1 & 1 & 0 & 0\\
i & -i & 0 & 0\\
0 & 0 & 1 & 1\\
0 & 0 & i & -i
\end{array}\right)
\end{equation*}
 is the transform matrix from $\alpha$ to $\Gamma$, satisfying $\Gamma=K\alpha$.
Therefore $\rho = \ee^{-\frac{1}{2}\alpha^T K^T {M^{-1}} K N_0 K^T {M^{-1}}^T K \alpha}$.
Comparing  this with Eq.\eqref{eq:rho-N} we get:
\begin{equation}
N_\phi=K^T M_{\phi}^{-1} K N_0 K^T {M_\phi^{-1}}^T K.
\end{equation}
With $N_\phi$ and Theorem \ref{theorem_fidelity} we can compute the Bures fidelity and finally the QFI for the lossy channel, which is exactly Eq.\eqref{equ13}.

\section{Calculation method of Eq. (\ref{equ14})}\label{append1}
Instead of the characteristic function description of density matrices, which is only convenient for the evaluation of the first moment of photon number operator, here we make use of the exponential quadratic operator and linear quantum transformation method \cite{Wang3} to evaluate the expectation value of observables.
According to its definitions, the parity operator is
$\hat{\Pi} = (-1)^{\hat{a}_1^\dag \hat{a}_1}=\ee^{-\ii\pi \hat{a}_1^\dag \hat{a}_1}.$
For our purpose, the evaluation of the parity operator's expectation value on the output state  $\braket{\hat{\Pi}}_{\phi}=\braket{\Psi_f|\hat{\Pi}|\Psi_f}$ can be formulated alternatively as \[\braket{\hat{\Pi}}_{\phi}=\braket{0|\hat{U}_{all}|0},\] where
\begin{equation}\label{all}
\begin{split}
&\hat{U}_{all}=\hat{S}^\dag\hat{U}_{lossyMZI}^\dag\hat{\Pi}\hat{U}_{lossyMZI}\hat{S},\\
&\hat{U}_{lossyMZI}=\hat{U}_{BS}\hat{U}_{loss}\hat{U}_{\phi}\hat{U}_{BS}.
\end{split}
\end{equation}
Here $\hat{S}=\exp{(-\zeta^*\hat{a}_1\hat{a}_2+\zeta\hat{a}_1^\dag\hat{a}_2^\dag)}$ is the two-mode squeeze operator and $\ket{0}$ denotes the vacuum state of the large system including the MZI and dissipative environment. Thus, our task transmits to calculating the expectation value of operator $\hat{U}_{all}$ over vacuum. This can be done by directly using the conclusion in \cite{Wang3}, stated as Theorem \ref{theorem_determinant} below.
\begin{theorem} \label{theorem_determinant}
Given an $n$-mode exponential quadratic operator $\hat{U}$, we define its matrix representation $M(\hat{U})$ by
$\hat{U} \Lambda^T \hat{U}^{-1} = \Lambda^T M(\hat{U})$. Here $\Lambda=\begin{pmatrix} \hat c^\dag \,, \hat c \end{pmatrix}^\text{T}$ with $\hat c^\dag = \begin{pmatrix} \hat{c}_1^\dag \,,\cdots\,, \hat{c}_n^\dag \end{pmatrix} \,, \hat c = \begin{pmatrix} \hat{c}_1 \,,\cdots\,, \hat{c}_n \end{pmatrix}$ and $[\hat{c}_i,\hat{c}_j^\dag]=\delta_{ij}$ holds.
Then the expectation value $\langle 0 | \hat{U} | 0 \rangle = \frac{1}{\sqrt {\det C}}$, if
 $M(\hat{U})=\left(\begin{array}{cc}A & D\\B & C\end{array}\right)$, where $A,B,C,D$ are all $n\times n$ matrices.
\end{theorem}
In our case, there are four modes in all, including  the two MZI modes $a_1,a_2$ and the two concomitant modes $e_1,e_2$ of the dissipative environment. Accordingly, here $\Lambda^T =(\hat{e}_1^\dag \,, \hat{e}_2^\dag \,, \hat{a}_1^\dag \,, \hat{a}_2^\dag \,, \hat{e}_1 \,, \hat{e}_2 \,, \hat{a}_1 \,, \hat{a}_2)$.
To apply Theorem 3, we only need to find out the matrix representation of operator $\hat{U}_{all}$ as defined by Eq. (\ref{all}).
It is easy to see
%\begin{widetext}
\begin{eqnarray}
\begin{split}
&M(\hat U_{all}) = M(\hat{S}^\dag)M(\hat{U}_{lossyMZI}^\dag)\\
&\times M(\hat{\Pi})M(\hat{U}_{lossyMZI})M(\hat{S}), \\
&M(\hat{U}_{lossyMZI}) = M(\hat{U}_{BS}) M(\hat{U}_{loss}) M(\hat{U}_{\phi}) M(\hat{U}_{BS}).
\end{split}
\end{eqnarray}
%\end{widetext}
Due to calculation detail, an unimportant global phase factor may appear. Here we give the $M$ matrices we made use of, where $\theta_i=\cos^{-1}(\sqrt{\eta_i})$ for $i=1,2$.
\begin{widetext}
\begin{eqnarray*}
M(\hat{\Pi})=\left(
\begin{array}{cccccccc}
 1 & 0 & 0 & 0 & 0 & 0 & 0 & 0 \\
 0 & 1 & 0 & 0 & 0 & 0 & 0 & 0 \\
 0 & 0 & -1 & 0 & 0 & 0 & 0 & 0 \\
 0 & 0 & 0 & 1 & 0 & 0 & 0 & 0 \\
 0 & 0 & 0 & 0 & 1 & 0 & 0 & 0 \\
 0 & 0 & 0 & 0 & 0 & 1 & 0 & 0 \\
 0 & 0 & 0 & 0 & 0 & 0 & -1 & 0 \\
 0 & 0 & 0 & 0 & 0 & 0 & 0 & 1 \\
\end{array}
\right),
M(\hat{S}) = |\zeta| \left(
\begin{array}{cccccccc}
 1 & 0 & 0 & 0 & 0 & 0 & 0 & 0 \\
 0 & 1 & 0 & 0 & 0 & 0 & 0 & 0 \\
 0 & 0 & \cosh (|\zeta|) & 0 & 0 & 0 & 0 & -\sinh (|\zeta|) \\
 0 & 0 & 0 & \cosh (|\zeta|) & 0 & 0 & -\sinh (|\zeta|) & 0 \\
 0 & 0 & 0 & 0 & 1 & 0 & 0 & 0 \\
 0 & 0 & 0 & 0 & 0 & 1 & 0 & 0 \\
 0 & 0 & 0 & -\sinh (|\zeta|) & 0 & 0 & \cosh (|\zeta|) & 0 \\
 0 & 0 & -\sinh (|\zeta|) & 0 & 0 & 0 & 0 & \cosh (|\zeta|) \\
\end{array}
\right),
\end{eqnarray*}
%\begin{widetext}
\begin{eqnarray*}
M(\hat{U}_{\phi}) = \left(
\begin{array}{cccccccc}
 1 & 0 & 0 & 0 & 0 & 0 & 0 & 0 \\
 0 & 1 & 0 & 0 & 0 & 0 & 0 & 0 \\
 0 & 0 & \ee^{-\ii \phi } & 0 & 0 & 0 & 0 & 0 \\
 0 & 0 & 0 & 1 & 0 & 0 & 0 & 0 \\
 0 & 0 & 0 & 0 & 1 & 0 & 0 & 0 \\
 0 & 0 & 0 & 0 & 0 & 1 & 0 & 0 \\
 0 & 0 & 0 & 0 & 0 & 0 & \ee^{\ii \phi } & 0 \\
 0 & 0 & 0 & 0 & 0 & 0 & 0 & 1 \\
\end{array}
\right),%\end{eqnarray*}
%\begin{eqnarray*}
 M(\hat{U}_{BS})=\left(
\begin{array}{cccccccc}
 1 & 0 & 0 & 0 & 0 & 0 & 0 & 0 \\
 0 & 1 & 0 & 0 & 0 & 0 & 0 & 0 \\
 0 & 0 & \frac{1}{\sqrt{2}} & \frac{\ii}{\sqrt{2}} & 0 & 0 & 0 & 0 \\
 0 & 0 & \frac{\ii}{\sqrt{2}} & \frac{1}{\sqrt{2}} & 0 & 0 & 0 & 0 \\
 0 & 0 & 0 & 0 & 1 & 0 & 0 & 0 \\
 0 & 0 & 0 & 0 & 0 & 1 & 0 & 0 \\
 0 & 0 & 0 & 0 & 0 & 0 & \frac{1}{\sqrt{2}} & -\frac{\ii}{\sqrt{2}} \\
 0 & 0 & 0 & 0 & 0 & 0 & -\frac{\ii}{\sqrt{2}} & \frac{1}{\sqrt{2}} \\
\end{array}
\right),\end{eqnarray*}
%\begin{eqnarray*}%\end{eqnarray*}
\begin{eqnarray*}
M(\hat{U}_{loss}) =\left(
\begin{array}{cccccccc}
 \cos \left(\theta _1\right) & 0 & \ii \sin \left(\theta _1\right) & 0 & 0 & 0 & 0 & 0 \\
 0 & \cos \left(\theta _2\right) & 0 & \ii \sin \left(\theta _2\right) & 0 & 0 & 0 & 0 \\
 \ii \sin \left(\theta _1\right) & 0 & \cos \left(\theta _1\right) & 0 & 0 & 0 & 0 & 0 \\
 0 & \ii \sin \left(\theta _2\right) & 0 & \cos \left(\theta _2\right) & 0 & 0 & 0 & 0 \\
 0 & 0 & 0 & 0 & \cos \left(\theta _1\right) & 0 & -\ii \sin \left(\theta _1\right) & 0 \\
 0 & 0 & 0 & 0 & 0 & \cos \left(\theta _2\right) & 0 & -\ii \sin \left(\theta _2\right) \\
 0 & 0 & 0 & 0 & -\ii \sin \left(\theta _1\right) & 0 & \cos \left(\theta _1\right) & 0 \\
 0 & 0 & 0 & 0 & 0 & -\ii \sin \left(\theta _2\right) & 0 & \cos \left(\theta _2\right) \\
\end{array}
\right).\end{eqnarray*}
\end{widetext}
With these, evaluation of the expectation value is straightforward; the result is given by Eq. (\ref{equ14}).


\begin{thebibliography}{16}
\bibitem[12]{upper bound} A physically perspicuous upper bound can be obtained using the QFI of the larger system (system considered plus the environment): $\tilde{F}_Q\equiv F_Q^{(S+E)}=\braket{\psi_f|\Delta(\hat{a}^\dag\hat{a})|\psi_f}=\eta^2n(n+2)+\eta(1-\eta)n$ with $\ket{\psi_f}$ being the TMSV state after photon loss for instance.
\end{thebibliography}
\end{document}